\documentclass[journal]{IEEEtran}
\ifCLASSINFOpdf
   \usepackage[pdftex]{graphicx}
   \graphicspath{{../pdf/}{../jpeg/}}
   \DeclareGraphicsExtensions{.pdf,.jpeg,.png,.JPG}
\else
\fi
\usepackage{amsmath}
\usepackage{amsfonts}
\usepackage{hyperref}
\begin{document}
%
\title{Data-Driven Continuum Dynamics via Transport-Teleport Duality}
%
%
%

\author{Jong-Hoon~Ahn \\
Nokia Bell Labs, Murrary Hill, NJ}

\maketitle

\begin{abstract}
In recent years, machine learning methods have been widely used to study physical systems that are challenging to solve with governing equations. Physicists and engineers are framing the data-driven paradigm as an alternative approach to physical sciences. In this paradigm change, the deep learning approach is playing a pivotal role. However, most learning architectures do not inherently incorporate conservation laws in the form of continuity equations, and they require dense data to learn the dynamics of conserved quantities. In this study, we introduce a clever mathematical transform to represent the classical dynamics as a point-wise process of disappearance and reappearance of a quantity, which dramatically reduces model complexity and training data for machine learning of transport phenomena. We demonstrate that just a few observational data and a simple learning model can be enough to learn the dynamics of real-world objects. The approach does not require the explicit use of governing equations and only depends on observation data. Because the continuity equation is a general equation that any conserved quantity should obey, the applicability may range from physical to social and medical sciences or any field where data are conserved quantities.
\end{abstract}

\begin{IEEEkeywords}
Physics-inspired video processing,
Physics-exploring machine learning, Data-driven physics, Representation learning.
\end{IEEEkeywords}

%
\IEEEpeerreviewmaketitle

\section{Introduction}
%
%
%
%
\IEEEPARstart{T}{he} real world is dynamic – all objects move, and environments change. On a daily life scale, the world is deterministic and predictable by the laws of physics. Solving governing equations is a firm and definite approach to study a physical system. However, the complexity of the real world often does not permit getting accurate and precise measurements or governing equations. 

As the methodology of machine learning has attracted significant attention from general science and engineering areas, physicists and engineers are getting increasingly interested in implementing the approach~\cite{Carleo2019,Brunton2020} for readily solving the governing equations~\cite{Tompson2017,Bezenac2019,Kim2019}, developing data-driven paradigms~\cite{Lusch2018,Reichstein2019,Yeo2019,Montans2019}, and discovering unknown governing equations~\cite{Raissi2018,Champion2019,Qin2019}.

The fundamental questions from physics society~\cite{Breen2019,Iten2020} include
\begin{itemize}
    \item How can machines understand human physics or perceive our physical world?
    \item Is deep learning about to reinvent physics?
    \item Could AI be an essential or general tool to explore new physics in the future?
\end{itemize}
Despite its considerable success in physics and the high expectations from the media and the general public, there are still many skeptics from physicists and expert groups~\cite{Montavon, Bathaee2017,Radovic2018,Buchanan2019,Tyagi2020}. Main issues frequently pointed out include
\begin{itemize}
    \item Deep learning requires highly complicated models with several tens of layers that are not physically well-interpreted,
    \item Deep learning models need exceedingly many training data that should mostly depend on human knowledge (governing equations) or other external simulation results (by solving the governing equations),
    \item Deep learning algorithms mostly work like {\it black boxes} that provide very little information about how they reached a certain conclusion.
\end{itemize}

In this study, we show research results that answer those fundamental questions and propose a framework for {\it physics-exploring machine learning} to settle out the issues. We suggest that those limitations of the existing deep learning models originate from the problem of representation~\cite{Zhong2016,Bengio2013}, and that a physical system or physical data should be re-represented before being applied to them. In short, we introduce how to turn the language of classical physics into a language for machine learning. Through the representation switching, we can dramatically reduce the complexity of existing machine learning models and the amount of training data required. We argue that understanding such duality of representation will contribute significantly to the development of machine learning theory for physical sciences in the future.

The main part of this paper is organized into two sections. In Section \ref{sec:DataDriven}, we first show our results and highlight the methods we applied. Then, in Section \ref{sec:duality}, we derive the detailed methods. \\

\begin{figure}[htp]
    \centering
    \includegraphics[width=8.2cm]{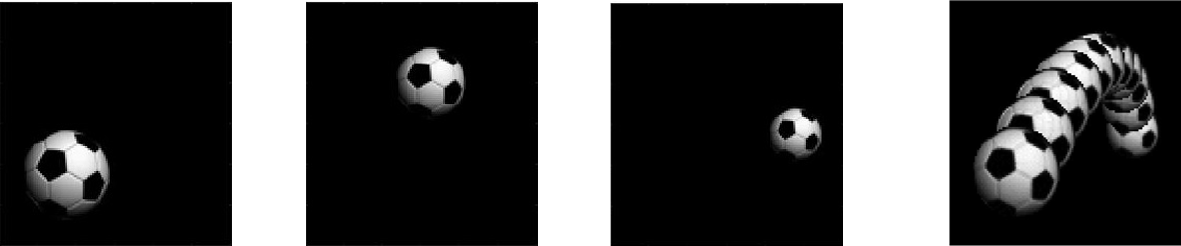}
    \caption{{\bf Generating a physical flow from sparse observations}. Three observation images of 128-by-128 pixels are treated as a mass distribution forming the shape of a soccer ball. The trace image is the output of our method from the three observations. By a physical flow, we mean a continuous flow of mass obeying the continuity equation.}
    \label{fig:soccerball}
\end{figure}

\section{Data-Driven Continuum Dynamics} \label{sec:DataDriven}
\subsection{Learning physics from observational data}
Consider the soccer ball example shown in Fig. \ref{fig:soccerball}. The soccer ball is not a point mass but a distributed mass; therefore, its motion cannot be modeled by a conventional interpolation of the coordinates of a point mass, or by the manifold learning~\cite{Roweis2000} of the pixel values from only three distributed mass images. Nevertheless, our method generates the intermediate and future ball images, following a curved trajectory as if it were affected by the gravitational field, from only three still images with no other pre-training data and no governing equation. Meanwhile, to generate images of an object in motion, a deep learning model is first trained on many example data to learn the effects of the same gravitational field on other objects, or governing equations. Our method is different from many physics-learned simulators or a physics-informed deep learning approach~\cite{Tompson2017,Kim2019,Raissi2018,Qin2019,Li2018,Umme2020,Sanchez20}. Our approach does not depend on human knowledge and enables machines to explore data of physical world by themselves.

\begin{figure}[htp]
    \centering
    \includegraphics[width=8.8cm]{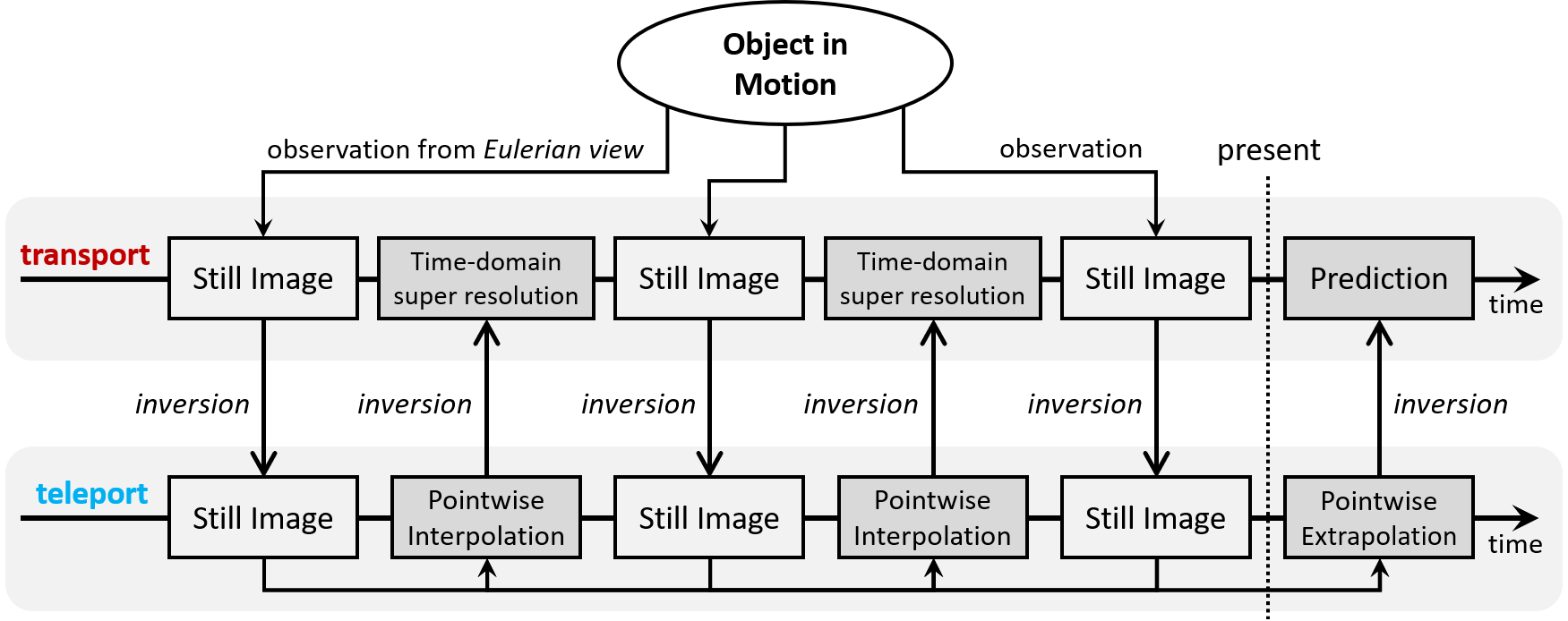}
    \caption{{\bf Flowchart}. When observing from the Eulerian view, we can obtain pixel images of a moving object. To generate a continuous physical flow from discrete time observations, we first transform them to another representation via an involution. In the new representation, object motions follow point-wise dynamics, which allows us to interpolate or extrapolate the images pixel-by-pixel. Finally, when we transform the interpolated or extrapolated images back to the original representation via the involution, we can obtain time domain super-resolution images or prediction images, which form a physical flow similar to that generated in Fig. \ref{fig:soccerball}.}
    \label{fig:flowchart}
\end{figure}

\begin{figure}[htp]
    \centering
    \includegraphics[width=8.2cm]{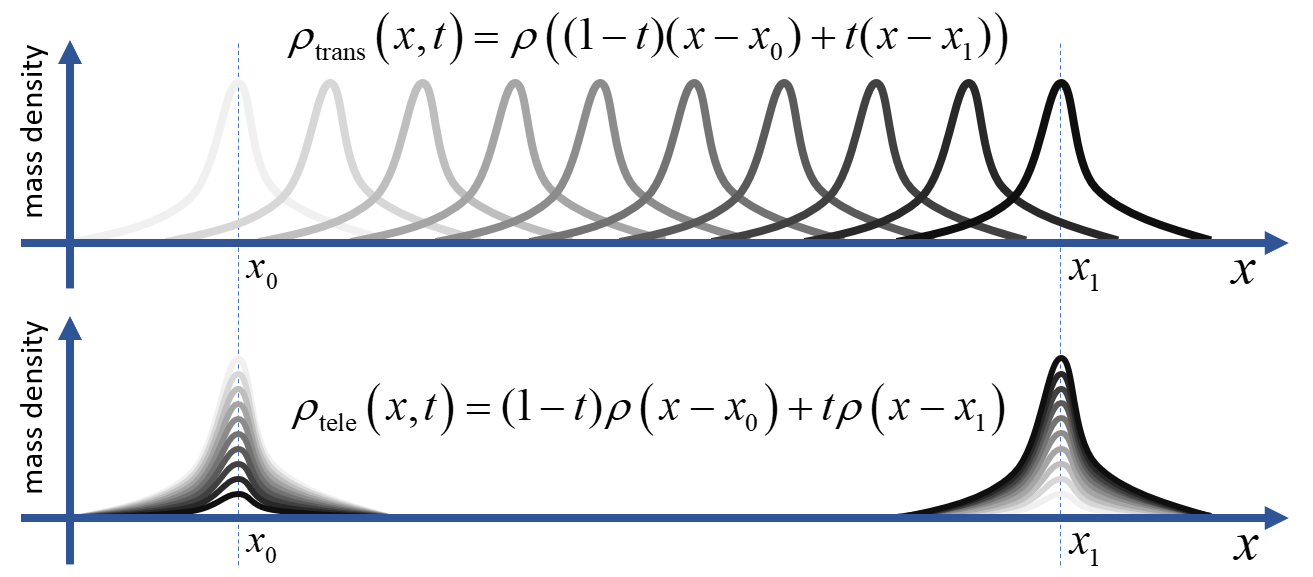}
    \caption{{\bf Example of mass transport and mass teleport}. The lower figure depicts a mass teleport. It resembles the hypothetical transfer of matter or energy from one point to another without traversing the physical space between them.}
    \label{fig:trans_tele}
\end{figure}

Our method is analogous to the numerical interpolation of data points, as illustrated in Fig. \ref{fig:flowchart}. Before interpolating them, however, we change each observation into another representation by inversion transform. That inversion transform is an involution that can change mass transport into a {\it mass teleport}, and then change back to the original transport by the same transform. The simplest case of a mass teleport is the linear combination of initial and final mass densities, as shown in Fig. \ref{fig:trans_tele} (The upper and lower figures of Fig. \ref{fig:trans_tele} are not direct transforms to each other). The soccer ball example of Fig. \ref{fig:soccerball} was generated by the second-order Lagrange interpolation of the three inverted images.

Transforming to a mass teleport is a velocity-free formulation of the continuum dynamics, in which we freeze the flow and let it transfer through the point-wise disappearance and reappearance of mass. Common complications in science, engineering, and machine learning originate from velocity. Solving the Navier-Stokes equations, determining the optical flow~\cite{Horn1981}, or tracking moving objects from video frames have been challenging problems regarding velocity. Our strategy is utilizing a much simpler machine learning model with fewer data after transforming the continuum dynamics into a velocity-free representation. Discovering a proper representation has been a fundamental issue in computer vision~\cite{Beymer1996,Agarwal2004,Wright2010}, computational neuroscience~\cite{Marr1982,Olshausen1996,DiCarlo2007}, and machine learning~\cite{Bengio2013,Roweis2000,Lee1999}.

\subsection{Inversion Transform}
\begin{figure}[htp]
    \centering
    \includegraphics[width=8.6cm]{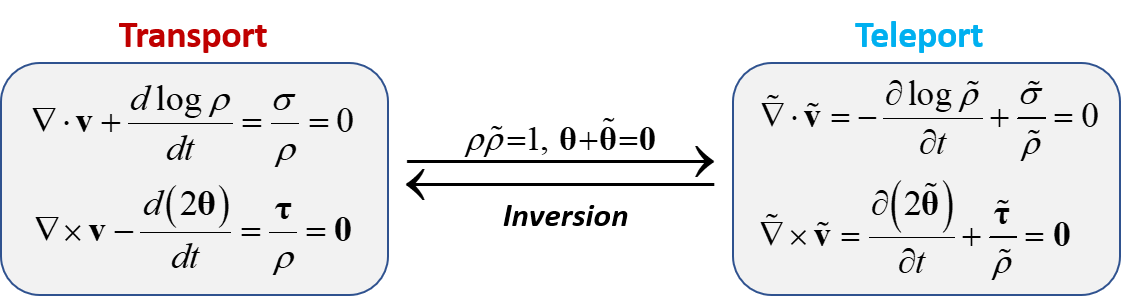}
    \caption{{\bf Representation switching of the continuum dynamics via inversion transform}. The inversion transform mentioned in Fig. \ref{fig:flowchart} is given by the reciprocal of the mass density and the negative of the rotational angles. The combination of two simple transforms can change the appearance of continuum dynamics, i.e., from transport to teleport, and from teleport back to transport.}
\label{fig:highlight}
\end{figure}

Figure \ref{fig:highlight} elaborates on what the inversion transform is and summarizes the highlights of our derivations, using which we can develop learning algorithms of observation data. In the case of mass transport, the movement of a physical object obeys the continuity equation (first equation on the upper left of Fig. \ref{fig:highlight}). However, the field of velocity is completely described by considering both its rotational and irrotational components. Thus, we need the second equation to determine the rotational components of motion. It states that the vorticity is twice the angular velocity at any point in a moving fluid~\cite{Lai2014}. The two equations impose fundamental constraints to conserved systems.

To discuss it in a more general way, we include the source or sink terms, $\sigma$ and $\boldsymbol{\tau}$, with double equalities. Then the first equation can be referred to as the general continuity equation. The local rate $\sigma$ is zero only when the mass density is conserved at a local point. In similar, we can define a vector of the local rates, $\boldsymbol{\tau}$, which is related to virtual sources or sinks to the local angles but does not make additional real rotations.

\begin{figure*}[htp]
    \centering
    \includegraphics[width=17.6cm]{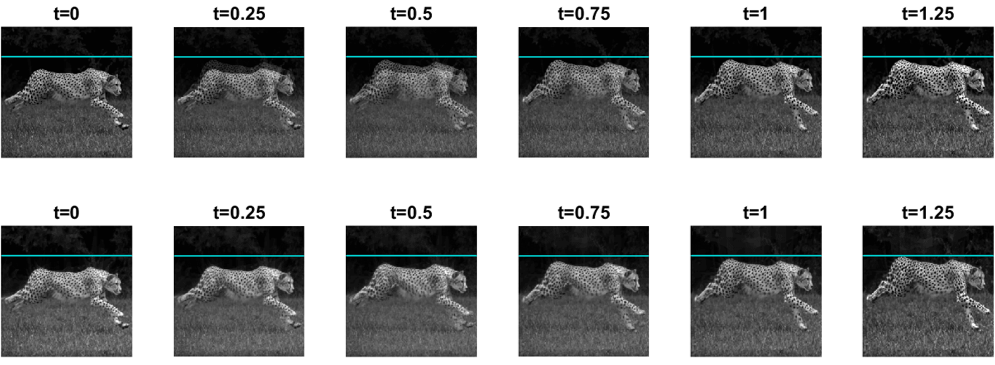}
    \caption{From two still images taken at time $t=0$ and $t=1$, we generated the three intermediate frames at $t = 0.25, 0.5, 0.75$ and one future frame at $t = 1.25$ by using the pixel-wise linear interpolation ({\it upper}) and our method ({\it lower}, pixel-wise linear interpolation after inversion transform). Pixel-wise linear interpolation outputs a superimposed image of the two, whereas our method finds a physical flow satisfying the continuity equation (\href{https://drive.google.com/file/d/1pVs7o3VA-yU5x_niJCoeUAgyP1yEBdwV/view?usp=sharing}{click \underline{here} to see a video clip}).}
    \label{fig:cheetah}
\end{figure*}

We can transform the transport equations into another form by using an amazingly simple inversion rule. It is given by the reciprocal of mass density and the negative of local angles. More specifically, they can be given by
\begin{equation}
    \tilde{\rho}(\tilde{\mathbf{x}})\equiv\det[\tilde{\mathbf{J}}]=\det[\mathbf{J}]^{-1}\equiv\rho^{-1}(\mathbf{x}) \label{eq:invrule1}
\end{equation}
and
\begin{equation}
    \tilde{\boldsymbol{\theta}}(\tilde{\mathbf{x}})\equiv\int [d\tilde{\mathbf{J}}~\tilde{\mathbf{J}}^{-1}]_{\times}=-\int[\mathbf{J}^{-1}d\mathbf{J}]_{\times}\equiv-\boldsymbol{\theta}(\mathbf{x}) \label{eq:invrule2}
\end{equation}
where $\mathbf{J}$ and $\tilde{\mathbf{J}}\equiv\mathbf{J}^{-1}$ are the Jacobian matrices defining a transformation between two spaces denoted by $\mathbf{x}\in\boldsymbol{\Omega}\subset\mathbb{R}^3$ and $\tilde{\mathbf{x}}\in\tilde{\boldsymbol{\Omega}}\subset\mathbb{R}^3$, and $[\cdot]_{\times}$ is an operator defined in Eq. (\ref{eq:crossop1}) (See Section \ref{sec:angles} for detailed information). Then, the time-derivative terms of the transport equations can move to the right side, while the double equalities are fixed. They immediately lead to zero-divergence and zero-curl of the velocity field on the right of Fig. \ref{fig:highlight}, which means that the velocity field is trivially constant, or entirely zero by anchoring a reference point. The mass teleport equations comprise only velocity-independent partial derivatives and become point-wise operations of both mass density and the angles of rotation. The local rates $\tilde{\sigma}$ and $\tilde{\boldsymbol{\tau}}$ determine the change of inverted mass density $\tilde{\rho}$ and its local angles of rotation $\tilde{\boldsymbol{\theta}}$. The process of modeling the local rates from observational data is what an interpolating polynomial or, generally, a deep learning model can do. The integration of the local rates is zero, indicating that the total mass and angles remain unchanged.

\begin{figure}[htp]
    \centering
    \includegraphics[width=8.6cm]{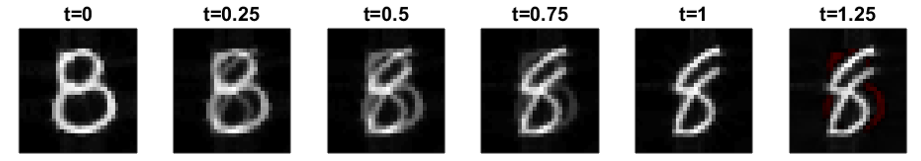}
    \includegraphics[width=8.6cm]{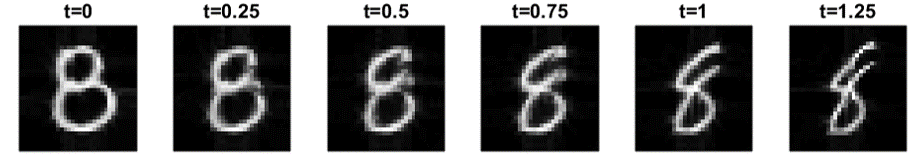}
    \caption{From two still images of a digit parameterized with $t=0$ and $t=1$, we generated the three intermediate frames at $t = 0.25, 0.5, 0.75$ and one extrapolated frame at $t = 1.25$ by using the pixel-wise linear interpolation ({\it upper}) and our method ({\it lower}, pixel-wise linear interpolation after inversion transform). Pixel-wise linear interpolation outputs a superimposed image of the two, whereas our method finds an underlying manifold onto which the two digit images can be embedded. The underlying manifold approaches a ground truth as the number of data increases.}
    \label{fig:digit}
\end{figure}

\subsection{Inversion of Incomplete Observations}
The existence of a zero-velocity representation of continuum dynamics provides us with a general framework for developing a model of real-world dynamics where there is no governing equation, as shown in Figs. \ref{fig:cheetah} and \ref{fig:digit}. The inverted dynamics is not coupled to its velocity field, and it is decomposed into point-wise dynamics of inverse mass. By learning the inverted point-wise dynamics from observational data, we can determine the dynamics of objects.

In Figs. \ref{fig:cheetah} and \ref{fig:digit}, we start with only two 256-by-256 pixel images of a running cheetah and two 32-by-32 pixel images of a handwritten digit, obtained at $t = 0$ and $t = 1$. To obtain an inversion of them without the local angles of rotation and the velocity field (many real-world problems fall into this case), we solved the following optimization problem:
\begin{equation}
    \min\sum_{t=0}^{T-1}\sum_{i,j}\left(\log\tilde{\rho}_{ij}^{(t)}-\log\tilde{\rho}_{ij}^{(t-1)}\right)^2\label{ErF}
\end{equation}
where $T = 1$ for the two-sample case and $(i, j)$ are the pixel indices. From Eq. (\ref{ErF}) we intend to minimize the local rates $\tilde{\rho}$ normalized by the density of inverted mass by the Euclidean metric. The operation is pixel-wise, and velocity information is not required. As shown in Fig. \ref{fig:flowchart}, once we obtain the inversion of original images, we can fit an interpolating function to them to generate the new frames at $t$ $=$ $0.25$, $0.5$, $0.75$, and $1.25$. Finally, by transforming them back to the original representation, we can reconstruct the frames of the running cheetah and a continuous variation of the digit. The generated frames do not necessarily correspond to the ground-truth of the cheetah motion and the ground-truth of the manifold~\cite{Roweis2000} onto which the digit images are embedded, but can approach them by increasing the frequency of observation. Nevertheless, we should note that the frames follow a physical flow that satisfies the continuity equation of mass. Eq. (\ref{ErF}) is one of the straightforward forms available and can be tailored into a more or less sophisticated form for better performance or practical use cases, respectively.

While many excellent algorithms have been proposed for video prediction from two-dimensional videos~\cite{Finn2016,Vondrick2016,Liu2017,Jin2017,Walker2016}, our method is promising for 3D volumetric data that provide complete information without occlusion. Two-dimensional natural images are not significant sources of conserved quantities. They usually comprise a two-layer structure of multiple foregrounds and background, and a foreground may occlude the background or the other foregrounds. The edges around those occlusions are non-zero sources or sinks of pixel values, which violate the conservation law. This explains why the left forefoot in the cheetah example appears a little distorted.

The algorithms for optical flow~\cite{Horn1981} can also be compared to our method. Both share the same goal of determining a flow from video frames. However, optical flow assumes only brightness constancy, whereas our method follows the strict form of continuity equation as an inviolable rule. Optical flow is suitable for visual sciences data, while our method is suitable for physical sciences data.

\section{Mass Transport-Teleport Duality} \label{sec:duality}

In this section, we show how the inversion transform changes the transport equations into the teleport equations. It is organized into thirteen subsections with sixty-six equations. In Section \ref{sec:teleport_inv}, the teleport equations are derived. In Section \ref{sec:inv_obs}, Eq. (\ref{ErF}) is derived. Section \ref{sec:intp} elaborates on interpolation and extrapolation.

\subsection{Transport of Mass} \label{sec:mass_transport}
By a physical flow that we mention in the paper, we mean that there is a flux of a quantity, and it should satisfy the continuity equation as a local conservation law. It is a partial differential equation which gives a relation between the amount of the quantity and the transport of that quantity. It states that the amount can only change by what is moved in or out. It is a universal equation that any transport of a non-interacting conserved quantity must obey~\cite{Lai2014}. In this paper, we focus on the discussion of mass and its transport.

The continuity equation has two forms depending on the viewpoint. From the Eulerian perspective, we have it in the form of
\begin{equation}
\nabla\cdot(\rho\mathbf{v})+\frac{\partial\rho}{\partial t}=0 \label{conEq1}
\end{equation}
where \(\rho\) is the density of mass. From the Lagrangian perspective, we have it in the form of
\begin{equation}
\nabla\cdot\mathbf{v}+\frac{d\log\rho}{dt}=0\label{conEq2}
\end{equation}
from Eq. (\ref{conEq1}) by
\[\nabla\cdot(\rho\mathbf{v})+\frac{\partial\rho}{\partial t}=\rho\nabla\cdot\mathbf{v}+(\mathbf{v}\cdot\nabla)\rho+\frac{\partial\rho}{\partial t}=\rho(\nabla\cdot\mathbf{v}+\frac{d\log\rho}{dt})\]
Eq. (\ref{conEq2}) describes a flux of the current of a unit mass density. Eq. (\ref{conEq1}) or Eq. (\ref{conEq2}) is a fundamental constraint that almost all physical quantities must obey as a local form of conservation laws.

In this paper, we introduce another quantity to describe the continuum dynamics. It is the local angle of rotation. It is a scalar for 2D. For 3D, it has the three components along x-, y-, and z-axis. In classical continuum dynamics, we are not interested in discussing the absolute values of a local angle of rotation, and we discuss vorticity more often than the local angular velocity. If we are given the velocity, the mass, and the pressure, the state of a fluid is completely determined. However, the purpose of our research work is to search for a new representation on which we can work on machine learning of physical data, and we want it to be a velocity-free formulation. Thus, we need to move our interest from the velocity to the angle of rotation. It begins with the equation of vorticity and angular velocity:
\begin{equation}
\nabla\times\mathbf{v}=2\frac{d\boldsymbol{\theta}}{dt}\label{conEq4}
\end{equation}
It is well-known that the vorticity is twice the angular velocity at any point in a moving fluid. In this paper, we explicitly denote the local angles of rotation in the equation. We interpret Eq. (\ref{conEq4}) as a complementary equation to Eq. (\ref{conEq2}). We refer to Eq. (\ref{conEq4}) as the second kind of continuity.

Eq. (\ref{conEq4}) has the total time derivative, which means that the local angles depend on the velocity. In Eq. (\ref{conEq3}), we will see a different version of Eq. (\ref{conEq4}). Eq. (\ref{conEq3}) is for the Eulerian description, and it will be complementary to Eq. (\ref{conEq1}). We will refer to Eqs. (\ref{conEq1}) and (\ref{conEq3}) as the equations of mass transport for the Eulerian view. We can also refer to Eqs. (\ref{conEq2}) and (\ref{conEq4}) as the equations of mass transport for the Lagrangian description. Now we are ready to explore a new representation of the continuum dynamics.

\subsection{Mass-Volume Conversion} \label{sec:conversion}
\begin{figure}[htp]
    \centering
    \includegraphics[width=8.6cm]{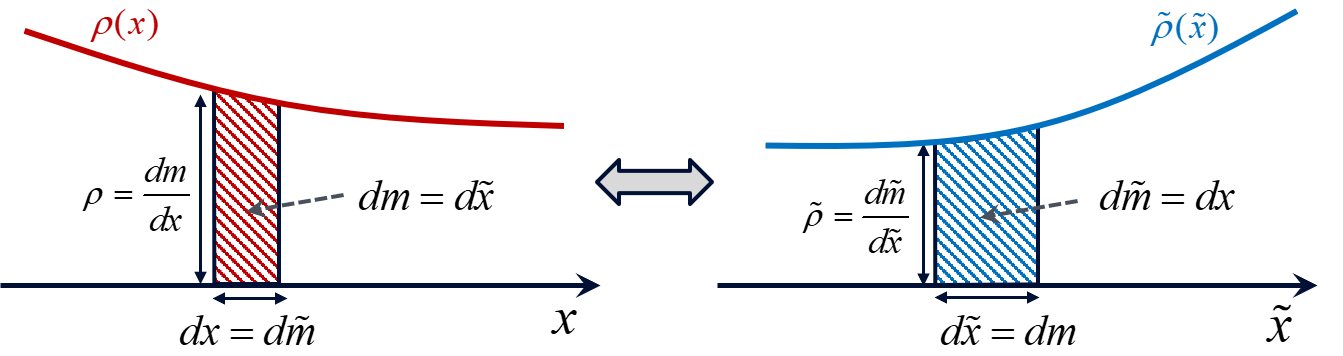}
    \caption{One-dimensional mass-volume conversion.}
    \label{fig:MVC}
\end{figure}

As a continuum, particles or material points of an object are represented by a mass density, which is conventionally mass per unit volume. In this section, what we try to propose is a mathematical process of exchanging mass elements with volume elements. By the process that we call mass-volume conversion for a one-dimensional case as shown in Fig. \ref{fig:MVC}, the mass density $\rho=dm/dx$ switches to its inverse mass density $\tilde{\rho}=dx/dm$. By introducing new notations for inverse mass and inverse volume, we can write it by the same definition of density, that is to say, $\tilde{\rho}=d\tilde{m}/d\tilde{x}$. For this, we need to assume that the volume element $dx$ can be converted into the new mass element $d\tilde{m}$, and the mass element $dm$ can be converted into the new volume element $d\tilde{x}$. The new mass density can switch back to the initial mass density by the same rules $d\tilde{m}=dx$ and $d\tilde{x}=dm$. It looks like a trivial transform, but it is not.

Let us continue to discuss it in a multi-dimensional space. As derived above, the one-dimensional mass density also becomes a one-dimensional volume ratio $\rho=dm/dx=d\tilde{x}/dx$  by converting the mass element into the new volume element. A multivariate analogue of the volume ratio $d\tilde{x}/dx$ is the Jacobian matrix $\left[\partial\tilde{\mathbf{x}}/\partial\mathbf{x}\right]$, where $\mathbf{x}\in\boldsymbol{\Omega}\subset\mathbb{R}^n$, $\tilde{\mathbf{x}}\in\tilde{\boldsymbol{\Omega}}\subset\mathbb{R}^n$.  To equate it to the scalar density $\rho$, we need to take the determinant operator to the Jacobian matrix. It is well consistent with the interpretation that the scalar ratio, $d\tilde{\mathbf{x}}/d\mathbf{x}$, is a volume ratio because the Jacobian determinant’s geometric meaning is the volumetric ratio. Thus, the mass-volume conversion in a multi-dimensional space leads to the equivalence of mass density and Jacobian determinant,
\begin{equation}
    \rho=\det\mathbf{J} \label{MVC1}
\end{equation}
where $\mathbf{J}=\mathbf{J}(\mathbf{x})=\left[\partial\tilde{\mathbf{x}}/\partial\mathbf{x}\right]$ is a positive definite Jacobian matrix and $\rho=\rho(\mathbf{x})$ is a positive density function. Similarly, we can have the same equivalence for the new mass density:
\begin{equation}
    \tilde{\rho}=\det\tilde{\mathbf{J}} \label{MVC2}
\end{equation}
where $\tilde{\mathbf{J}}=\tilde{\mathbf{J}}(\tilde{\mathbf{x}})=\left[\partial\mathbf{x}/\partial\tilde{\mathbf{x}}\right]$ is the inverse matrix of $\tilde{\mathbf{J}}$ and $\tilde{\rho}=\tilde{\rho}(\tilde{\mathbf{x}})$ is the reciprocal of the original function $\rho=\rho(\mathbf{x})$, that is,
\begin{equation}
    \tilde{\rho}=\frac{1}{\rho} \label{inversion1}
\end{equation}
We can refer to Eq. (\ref{inversion1}) as the mass-mass inversion rule that is shown in Figs. \ref{fig:highlight} and \ref{fig:Representation}. We can learn that the equivalence of mass density and Jacobian determinant is symmetric under the conversion process. It looks trivial again, but it is not a simple reciprocal operation because it involves a nonlinear coordinate transformation between $\mathbf{x}$ and $\tilde{\mathbf{x}}$. If either is the mass density, we can refer to the other as the density of inverse mass or inverted mass density. This inversion transform originates from Legendre transform~\cite{Zia2009}.

Eq. (\ref{MVC1}) can also be understood in Monge’s continuous formulation~\cite{Haker2004,Kolouri2017}. The optimal mass transportation problem is to find a mapping that transports a first mass distribution onto a second mass distribution at a minimal cost. The general formulation applies to our problem in two ways. First, the second mass distribution becomes uniform in Eq. (\ref{MVC1}). Second, we can also consider the opposite case, which is its inverse mapping from the uniform second distribution to the initial mass distribution, and can derive Eq. (\ref{MVC2}). Our purpose is to transform a mass distribution to its inverse distribution, while optimal mass transportation finds an optimal mapping between two mass distributions.

\subsection{Differential Forms}

The mass-volume conversion has the definite form by Eqs. (\ref{MVC1}) and (\ref{MVC2}), but it can be more powerful when we put it in a differential form that is independent of coordinates. By taking the logarithm and performing some calculus by
\begin{eqnarray*}
&&\rho=\det\mathbf{J} \nonumber \\
&\Rightarrow&\log\rho=\log\left[\det\mathbf{J}\right]=\mathrm{Tr}\left[\log\mathbf{J}\right] \\
&\Rightarrow&d\left(\log\rho\right)=d\left(\mathrm{Tr}\left[\log\mathbf{J}\right]\right)=\mathrm{Tr}\left[d\left(\log\mathbf{J}\right)\right]=\mathrm{Tr}\left[\mathbf{J}^{-1}d\mathbf{J}\right],
\end{eqnarray*}
we can get its differential form
\begin{equation}
    d\left(\log\rho\right)=\mathrm{Tr}\left[\mathbf{J}^{-1}d\mathbf{J}\right] \label{dMVC1}
\end{equation}
We can utilize Eq. (\ref{dMVC1}) in various ways. Examples include the following cases:
\begin{eqnarray}
\nabla\left(\log\rho\right)&=&\mathrm{Tr}[\mathbf{J}^{-1}\nabla\mathbf{J}] \label{dMVC1_1} \\
\frac{d\log\rho}{dt}&=&\mathrm{Tr}[\mathbf{J}^{-1}\frac{d\mathbf{J}}{dt}] \label{dMVC1_2} \\
\left(\mathbf{v}\cdot\nabla\right)\left(\log\rho\right)&=&\mathrm{Tr}[\mathbf{J}^{-1}\left(\mathbf{v}\cdot\nabla\right)\mathbf{J}]. \label{dMVC1_3} 
\end{eqnarray}
Similarly, we can also get a similar form from Eq. (\ref{MVC2}):
\begin{equation}
    d\left(\log\tilde{\rho}\right)=\mathrm{Tr}[d\tilde{\mathbf{J}}~\tilde{\mathbf{J}}^{-1}] \label{dMVC2}
\end{equation}
Generally, the differential Jacobian and the inverse Jacobian are not commutative, but we can keep it commutative inside the trace operator. We can utilize Eq. (\ref{dMVC2}) to have
\begin{eqnarray}
\tilde{\nabla}\left(\log\tilde{\rho}\right)&=&\mathrm{Tr}[\tilde{\nabla}\tilde{\mathbf{J}}~\tilde{\mathbf{J}}^{-1}] \label{dMVC2_1} \\
\frac{d\log\tilde{\rho}}{dt}&=&\mathrm{Tr}[\frac{d\tilde{\mathbf{J}}}{dt}\tilde{\mathbf{J}}^{-1}] \label{dMVC2_2}
\end{eqnarray}
where $\tilde{\nabla}$ is the gradient regarding the new coordinates $\tilde{\mathbf{x}}$ .

Because $\tilde{\mathbf{J}}$ is the inverse matrix of $\mathbf{J}$, we have the identity
\begin{equation}
    \mathbf{J}^{-1}d\mathbf{J}+d\tilde{\mathbf{J}}~\tilde{\mathbf{J}}^{-1}=\mathbf{O}, \label{idt1}
\end{equation}
where $\mathbf{O}$ is a matrix whose components are all zero. Then, we get differential forms of the mass-Jacobian conversion rules:
\begin{eqnarray}
d\left(\log\rho\right)+\mathrm{Tr}[d\tilde{\mathbf{J}}~\tilde{\mathbf{J}}^{-1}]=0, \label{dMVC3} \\
d\left(\log\tilde{\rho}\right)+\mathrm{Tr}[\mathbf{J}^{-1}d\mathbf{J}]=0. \label{dMVC4}
\end{eqnarray}

\subsection{Incorporating the Angles} \label{sec:angles}
The conversion of mass density and Jacobian determinant tells that the mass elements and volume elements are exchangeable, but the mass is a scalar quantity and Jacobian is a tensor quantity. We need extra dimensions to determine the Jacobian matrix. For a compact notation, we define a linear operator denoted by $[\cdot]_\times$. When we apply it to a $3\times 3$ matrix, the operator outputs a column vector by
\begin{equation}
\left[\begin{pmatrix}
A_{11} & A_{12} & A_{13} \\
A_{21} & A_{22} & A_{23} \\
A_{31} & A_{32} & A_{33}
\end{pmatrix}\right]_{\times} = \frac{1}{2}\begin{pmatrix}
A_{23}-A_{32} \\
A_{31}-A_{31} \\
A_{12}-A_{21} 
\end{pmatrix} \label{eq:crossop1}
\end{equation}
When it applies to a column, it can also output a matrix by
\begin{equation}
\left[\begin{pmatrix}
a_1 \\
a_2 \\
a_3
\end{pmatrix}\right]_{\times} = \begin{pmatrix}
0 & a_3 & -a_2 \\
-a_3 & 0 & a_1 \\
a_2 & -a_1 & 0
\end{pmatrix} \label{eq:crossop2}
\end{equation}
The operator is an involution for a scalar, a vector, and an anti-symmetric matrix. The Levi-Civita symbols can replace the first use case of the operator, but we want to perform both cases with a single compact notation.

Now, by applying the operator to the normalized differential Jacobian $\mathbf{J}^{-1}d\mathbf{J}$ that appears in Eq. (\ref{dMVC1}), we can make a connection to the local angles of rotation:
\begin{equation}
    d\boldsymbol{\theta} = \left[\mathbf{J}^{-1}d\mathbf{J}\right]_{\times} \label{dthJ1}
\end{equation}
The proof is straightforward, and we omit it. Similarly, we can derive
\begin{equation}
    d\tilde{\boldsymbol{\theta}} = [d\tilde{\mathbf{J}}~\tilde{\mathbf{J}}^{-1}]_{\times}. \label{dthJ2}
\end{equation}
By Eq. (\ref{idt1}), we have
\begin{equation}
    d\boldsymbol{\theta}+d\tilde{\boldsymbol{\theta}}=\mathbf{0} \label{idt2}
\end{equation}
where $\mathbf{0}$ is a column vector whose components are all zero. By Eq. (\ref{idt1}), we also have
\begin{eqnarray}
d\boldsymbol{\theta}+[d\tilde{\mathbf{J}}~\tilde{\mathbf{J}}^{-1}]_{\times}=\mathbf{0}, \label{dthJ3} \\
d\tilde{\boldsymbol{\theta}}+[\mathbf{J}^{-1}d\mathbf{J}]_{\times}=\mathbf{0}. \label{dthJ4}
\end{eqnarray}

Unfortunately, we are generally not allowed to have definite for the angles forms from integrating Eqs. (\ref{dthJ1}), (\ref{dthJ2}), (\ref{dthJ3}), or (\ref{dthJ4}) because the differential Jacobian and the inverse Jacobian are not commutative. If the matrices in the bracket can be commutative in a fixed eigenspace of the Jacobian, we may have definite forms of
\begin{eqnarray}
\boldsymbol{\theta}+[\log\tilde{\mathbf{J}}]_{\times}=\mathbf{0} \label{thJ1}, \\
\tilde{\boldsymbol{\theta}}+[\log\mathbf{J}]_{\times}=\mathbf{0}. \label{thJ2}
\end{eqnarray}
It is a particular case in which we can find a locally linear embedding of nonlinear dynamics.

\subsection{Transport of Jacobian} \label{sec:Jacobian_transport}
We usually consider a field of the Jacobian matrix for a curvilinear coordinate transform, and it is usually a static situation. However, the curvilinear coordinates we consider move and the Jacobian changes. In this subsection, we merge Eqs. (\ref{conEq2}) and (\ref{conEq4}) to derive a transport equation of Jacobian. Two equations in Eqs. (\ref{conEq2}) and (\ref{conEq4}) are complementary to each other. Before merging them, we need to consider the time-derivative form of Eq. (\ref{dthJ1}):
\begin{equation}
    \frac{d\boldsymbol{\theta}}{dt} = \left[\mathbf{J}^{-1}\frac{d\mathbf{J}}{dt}\right]_{\times}. \label{dtthJ1}
\end{equation}
First, by using Eq. (\ref{dMVC1_2}), we can factor out the trace operator from Eq. (\ref{conEq2}):
\begin{eqnarray}
0&=&\nabla\cdot\mathbf{v}+\frac{d\log\rho}{dt} \nonumber \\
&=&\mathrm{Tr}[\nabla\mathbf{v}]+\mathrm{Tr}[\mathbf{J}^{-1}\frac{d\mathbf{J}}{dt}] \nonumber \\
&=&\mathrm{Tr}[\nabla\mathbf{v}+\mathbf{J}^{-1}\frac{d\mathbf{J}}{dt}] \label{conEq2mer}
\end{eqnarray}
Second, by using Eq. (\ref{dtthJ1}), we can factor out the bracket operator from Eq. (\ref{conEq4}):
\begin{eqnarray}
0&=&\nabla\times\mathbf{v}-\frac{d(2\boldsymbol{\theta})}{dt} \nonumber \\
&=&-2[\nabla\mathbf{v}]_{\times}-2[\mathbf{J}^{-1}\frac{d\mathbf{J}}{dt}]_{\times} \nonumber \\
&=&-2[\nabla\mathbf{v}+\mathbf{J}^{-1}\frac{d\mathbf{J}}{dt}]_{\times} \label{conEq4mer}
\end{eqnarray}
Finally, we can get a merged equation from Eqs. (\ref{conEq2mer}) and (\ref{conEq4mer}):
\begin{equation}
\nabla\mathbf{v}+\mathbf{J}^{-1}\frac{d\mathbf{J}}{dt}=\mathbf{O} \label{transJac0}
\end{equation}
We can further transform Eq. (\ref{transJac0}) into what we call the transport equation of Jacobian
\begin{equation}
\nabla\left(\mathbf{Jv}\right)+\frac{\partial\mathbf{J}}{\partial t}=\mathbf{O} \label{transJac}
\end{equation}
by performing the following calculus with Eq. (\ref{transJac0}):
\begin{eqnarray*}
\nabla\left(\mathbf{Jv}\right)+\frac{\partial\mathbf{J}}{\partial t}&=&\mathbf{J}(\nabla\mathbf{v})+(\nabla\mathbf{J})\mathbf{v}+\frac{\partial\mathbf{J}}{\partial t} \\
&=&\mathbf{J}\left(\nabla\mathbf{v}\right)+(\mathbf{v}\cdot\nabla)\mathbf{J}+\frac{\partial\mathbf{J}}{\partial t} \\
&=&\mathbf{J}(\nabla\mathbf{v})+\frac{d\mathbf{J}}{dt} \\
&=&\mathbf{J}(\nabla\mathbf{v}+\mathbf{J}^{-1}\frac{d\mathbf{J}}{dt})=\mathbf{O}
\end{eqnarray*}
For deriving Eq. (\ref{transJac}), we also used the following identity regarding the directional derivatives of Jacobian:
\begin{equation}
    \left(\mathbf{v}\cdot\nabla\right)\mathbf{J}=\left(\nabla\mathbf{J}\right)\mathbf{v}, \label{gradJv}
\end{equation}
which comes from
\begin{eqnarray*}
(\mathbf{v}\cdot\nabla)\mathbf{J}&=&(v_k\frac{\partial}{\partial x_k})J_{ij} \\
&=&v_k\frac{\partial}{\partial x_k}\left(\frac{\partial \tilde{x}_i}{\partial x_j}\right) \\
&=&\frac{\partial}{\partial x_j}\left(\frac{\partial \tilde{x}_i}{\partial x_k}\right) v_k\equiv\left(\nabla\mathbf{J}\right)\mathbf{v}.
\end{eqnarray*}
The transport equation of Jacobian in Eq. (\ref{transJac}) resembles the transport equation of mass in Eq. (\ref{conEq1}). Eqs. (\ref{conEq2}) and (\ref{conEq4}) can be replaced by Eq. (\ref{transJac}) for the Lagrangian description, as explained in Section \ref{sec:viewpoints}.

\subsection{The Velocity is Zero}
In Section \ref{sec:conversion}, we have derived the conjugated density of mass called the density of inverse mass through Eqs. (\ref{MVC2}) and (\ref{inversion1}). Now we get one of the most important results in the paper and discuss it. Let us find out what happens to the inverse mass when the original mass transports by Eq. (\ref{transJac}) or, equivalently, Eqs. (\ref{conEq2}) and (\ref{conEq4}). First, let us consider the gradient of the inverse mass’s velocity. We learn that the gradient of the velocity is always zero:
\begin{equation}
    \tilde{\nabla}\tilde{\mathbf{v}}=\mathbf{0} \label{zerov}
\end{equation}
from using Eq. (\ref{transJac}) by performing the following calculus
\begin{eqnarray}
\tilde{\nabla}\tilde{\mathbf{v}}&=&(\tilde{\mathbf{J}}^T\nabla)(\frac{d\tilde{\mathbf{x}}}{dt}) \nonumber \\
&=&(\tilde{\mathbf{J}}^T\nabla)[(\mathbf{v}\cdot\nabla)\tilde{\mathbf{x}}+\frac{\partial\tilde{\mathbf{x}}}{\partial t}] \nonumber \\
&=&[\nabla(\mathbf{Jv}+\frac{\partial\tilde{\mathbf{x}}}{\partial t})]\tilde{\mathbf{J}} \nonumber \\
&=&[\nabla\left(\mathbf{Jv}\right)+\frac{\partial\tilde{\mathbf{J}}}{\partial t}]\mathbf{J}^{-1} = \mathbf{O} \label{transJacDer}
\end{eqnarray}
It readily means zero velocity at all points, as long as we anchor the reference point, by
\begin{equation}
    \tilde{\mathbf{v}}=\int d\tilde{\mathbf{v}}=\int_{\text{ref. point}}^{\tilde{\mathbf{x}}}(\tilde{\nabla}\tilde{\mathbf{v}})~d\tilde{\mathbf{x}}=\mathbf{0} 
\end{equation}

The canonical way of object transfer in physics is transportation. However, Eq. (\ref{zerov}) suggests that the inverse mass can move in a totally different way. In science fiction, film, video game, or the virtual world, we often imagine the theoretical transfer of objects: teleportation. An object can disappear in a place and simultaneously reappear in another distant place without physically crossing the space in between. In the following subsections, we will show that the dynamics of inverse mass follows such a teleportation-like way, as illustrated in Fig. \ref{fig:trans_tele}.

\subsection{Teleport of Inverse Jacobian} \label{sec:invJacobian_teleport}
Let us define a tensor quantity denoted by $\boldsymbol{\Sigma}$ as
\begin{equation}
\boldsymbol{\Sigma}\equiv\nabla(\mathbf{Jv})+\frac{\partial\mathbf{J}}{\partial t}. \label{Sigma}
\end{equation}
Then, we can simply express the transport equation of Jacobian of Eq. (\ref{transJac}) as $\boldsymbol{\Sigma}=\mathbf{O}$. For a symmetric formulation, we can also define
\begin{equation}
\tilde{\boldsymbol{\Sigma}}\equiv\tilde{\nabla}(\mathbf{\tilde{J}\tilde{v}})+\frac{\partial\tilde{\mathbf{J}}}{\partial t}. \label{tSigma}
\end{equation}
However, $\tilde{\boldsymbol{\Sigma}}$ is not a zero matrix. Let us learn what it is by performing the following calculus that is similar to Eq. (\ref{transJacDer}):
\begin{eqnarray}
\nabla\mathbf{v}&=&(\mathbf{J}^T\tilde{\nabla})(\frac{d\mathbf{x}}{dt}) \nonumber \\
&=&(\mathbf{J}^T\tilde{\nabla})[(\tilde{\mathbf{v}}\cdot\tilde{\nabla})\mathbf{x}+\frac{\partial\mathbf{x}}{\partial t}] \nonumber \\
&=&[\tilde{\nabla}(\mathbf{\tilde{J}\tilde{v}}+\frac{\partial\mathbf{x}}{\partial t})]\mathbf{J} \nonumber \\
&=&[\tilde{\nabla}(\mathbf{\tilde{J}\tilde{v}})+\frac{\partial\mathbf{J}}{\partial t}]\tilde{\mathbf{J}}^{-1} \label{transJacDer2}
\end{eqnarray}
which leads, by Eq. (\ref{tSigma}), to
\begin{equation}
    \nabla\mathbf{v}=\tilde{\boldsymbol{\Sigma}}\tilde{\mathbf{J}}^{-1} \label{gradv1}
\end{equation}

On the other hand, we have the following time-dependent version of Eq. (\ref{idt1}) by $\tilde{\mathbf{v}}=\mathbf{0}$:
\begin{equation}
    \mathbf{J}^{-1}\frac{d\mathbf{J}}{dt}+\frac{\partial\tilde{\mathbf{J}}}{\partial t}~\tilde{\mathbf{J}}^{-1}=\mathbf{O}. \label{tidt}
\end{equation}
We need to notice the partial derivative, which means that the velocity of inverse Jacobian is zero from Eq. (\ref{zerov}). From Eqs. (\ref{transJac0}), (\ref{gradv1}), and (\ref{tidt}), we finally get
\begin{equation}
    \frac{\partial\tilde{\mathbf{J}}}{\partial t}=\tilde{\boldsymbol{\Sigma}} \label{teleJ1}
\end{equation}
or, equivalently,
\begin{equation}
    \frac{\partial\tilde{\mathbf{J}}}{\partial t}\tilde{\mathbf{J}}^{-1}=\tilde{\boldsymbol{\Sigma}}\tilde{\mathbf{J}}^{-1}. \label{teleJ2}
\end{equation}
We can double-check it directly from Eq. (\ref{tSigma}) by replacing the velocity term with zero. Eq. (\ref{teleJ1}) does not have any velocity-dependent term, and the inverse Jacobian is governed by the local rate $\tilde{\boldsymbol{\Sigma}}$.

The local rate is not arbitrary. It should follow Eq. (\ref{gradv1}), which comes from the gradient of a known or observed velocity field. Even if we don’t know the velocity field, it leaves a constraint that the local rate should follow. Another form of Eq. (\ref{gradv1}) is $\tilde{\boldsymbol{\Sigma}}=\tilde{\nabla}\mathbf{v}$ . From the identity of $\tilde{\nabla}\times\tilde{\nabla}v_i=\mathbf{0}$ , we can learn that the curl of each row of $\tilde{\boldsymbol{\Sigma}}$ is always zero. We can simply denote it as the curl of a tensor
\begin{equation}
\tilde{\nabla}\times\tilde{\boldsymbol{\Sigma}}=\epsilon_{ijk}\tilde{\Sigma}_{mj,i}\mathbf{e}_k\otimes\mathbf{e}_m=\mathbf{O}
\end{equation}
By Eq. (\ref{teleJ1}), we can also make sure that the inverse Jacobian always has no circulation at any point:
\begin{equation}
\tilde{\nabla}\times\tilde{\mathbf{J}}=\mathbf{O} \label{zerocurlJ}
\end{equation}
which means $\oint_{any}\tilde{\mathbf{J}}d\tilde{\mathbf{l}}=\mathbf{0}$, and we can discretize Eq. (\ref{zerocurlJ}) for 2D case into
\begin{equation}
(\tilde{\mathbf{J}}_{ij}-\tilde{\mathbf{J}}_{i+1,j+1})\begin{pmatrix}
1 \\ -1\end{pmatrix}
=(\tilde{\mathbf{J}}_{i,j+1}-\tilde{\mathbf{J}}_{i+1,j})\begin{pmatrix}
1 \\ 1\end{pmatrix}. \label{nocurlJ2D}
\end{equation}
Because a field of Jacobian defines a curvilinear coordinate system, it is not allowed to have a circulation. We can refer to Eqs. (\ref{teleJ1}) and (\ref{zerocurlJ}) as the teleport equations of inverse Jacobian.

\subsection{Advective Rotation}
The advection operator is $\mathbf{v}\cdot\nabla$, and the advection of local angles is $(\mathbf{v}\cdot\nabla)\boldsymbol{\theta}$. It can be changed into another form by taking the following steps with the Levi-Civita symbol from Eq. (\ref{dtthJ1}): 
\begin{eqnarray*}
\frac{d\mathbf{\theta}}{dt}&=&[\mathbf{J}^{-1}\frac{d\mathbf{J}}{dt}]_{\times} \\
&=&[\mathbf{J}^{-1}((\mathbf{v}\cdot\nabla)\mathbf{J}+\frac{\partial\mathbf{J}}{\partial t})]_{\times} \\
&=&[\mathbf{J}^{-1}(\nabla\mathbf{J})\mathbf{v}]_{\times}+[\mathbf{J}^{-1}\frac{\partial\mathbf{J}}{\partial t}]_{\times} \\
&=&\frac{1}{2}\epsilon_{ijk}J_{jl}^{-1}(\partial_kJ_{lm})v_m+\frac{\partial\boldsymbol{\theta}}{\partial t} \\
&=&\frac{1}{2}\epsilon_{ijk}J_{ml}^{-1}(\partial_kJ_{lm})v_j+\frac{\partial\boldsymbol{\theta}}{\partial t} \\
&=&\frac{1}{2}\epsilon_{ijk}v_j\partial_k(\log\rho)+\frac{\partial\boldsymbol{\theta}}{\partial t} \\
&=&\frac{1}{2}\mathbf{v}\times(\nabla\log\rho)+\frac{\partial\boldsymbol{\theta}}{\partial t}
\end{eqnarray*}
where Eqs. (\ref{dMVC1_1}) and (\ref{gradJv}) are necessary for derivation. When we compare it to the chain rule $\frac{d\boldsymbol{\theta}}{dt}=(\mathbf{v}\cdot\nabla)\boldsymbol{\theta}+\frac{\partial\boldsymbol{\theta}}{\partial t}$, we get 
\begin{equation}
\mathbf{v}\times\nabla\log\rho=(\mathbf{v}\cdot\nabla)(2\boldsymbol{\theta}) \label{advrot1}
\end{equation}
or, equivalently,
\begin{equation}
\nabla\log\rho\times\mathbf{v}+\nabla(2\boldsymbol{\theta})\mathbf{v}=\mathbf{0} \label{advrot2}
\end{equation}
We can also derive Eqs. (\ref{advrot1}) or (\ref{advrot2}) directly from Eq. (\ref{gradJv}) by multiplying the inverse Jacobian to the left side of Eq. (\ref{gradJv}) and then taking the cross operator $[\cdot]_{\times}$ .

\begin{figure}[htp]
    \centering
    \includegraphics[width=6cm]{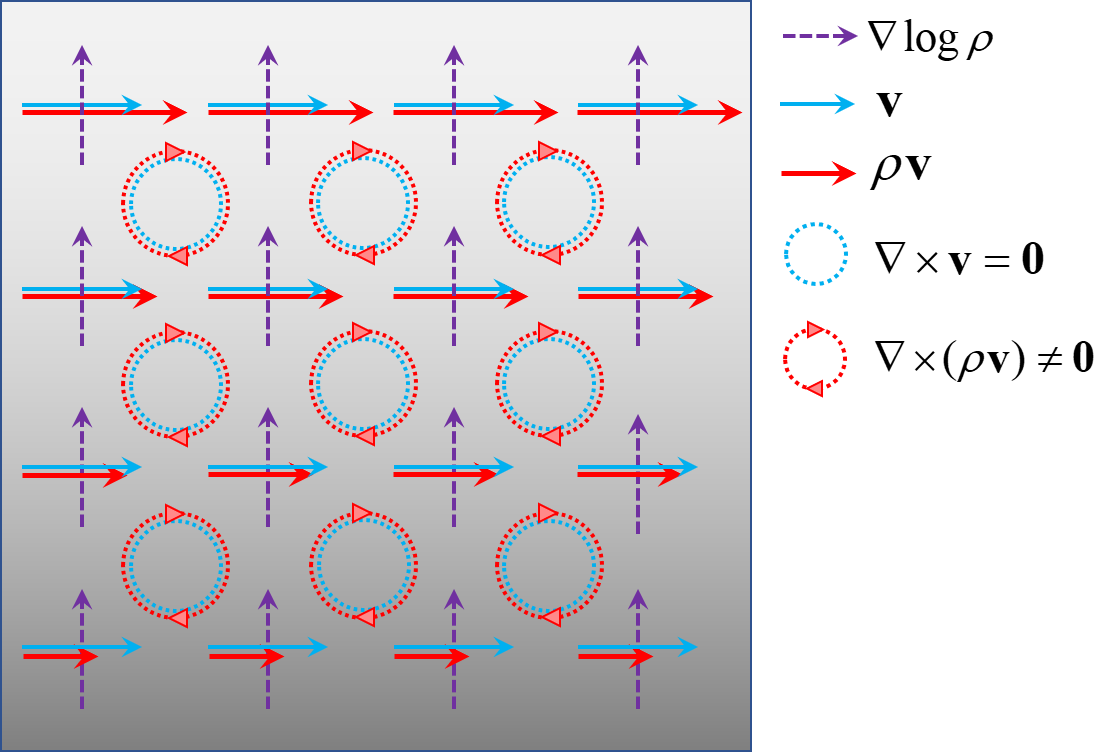}
    \caption{An example of zero vorticity. The mass current has non-zero curl despite zero vorticity.}
    \label{fig:zerovorticity}
\end{figure}

\subsection{Complete Eulerian Form}
The reason why we derive the equation of advective rotation is that we can get the Eulerian form of Eq. (\ref{conEq4}). It is important for us to derive the teleport equations of inverse mass in the next subsection. From Eqs. (\ref{conEq4}) and (\ref{advrot1}), we have
\begin{eqnarray*}
\mathbf{0}&=&\nabla\times\mathbf{v}-2\frac{d\boldsymbol{\theta}}{dt} \\
&=&\nabla\times\mathbf{v}-2(\mathbf{v}\cdot\nabla)\mathbf{\theta}-2\frac{\partial\boldsymbol{\theta}}{\partial t} \\
&=&\nabla\times\mathbf{v}-(\mathbf{v}\times\nabla)\log\rho-2\frac{\partial\boldsymbol{\theta}}{\partial t} \\
&=&\frac{1}{\rho}(\rho\nabla\times\mathbf{v}+\nabla\rho\times\mathbf{v}-2\rho\frac{\partial\boldsymbol{\theta}}{\partial t}) \\
&=&\frac{1}{\rho}(\nabla\times(\rho\mathbf{v})-2\rho\frac{\partial\boldsymbol{\theta}}{\partial t})
\end{eqnarray*}
which gives
\begin{equation}
\nabla\times(\rho\mathbf{v})-2\rho\frac{\partial\boldsymbol{\theta}}{\partial t}=0 \label{conEq3}
\end{equation}
The continuity equation, Eq. (\ref{conEq1}), states how the mass density should increase or decrease at a fixed point when a mass current converges or diverges at the point. Likewise, Eq. (\ref{conEq3}) tells how the local angle of rotation should decrease or increase at a fixed point when a mass current rotates clockwise or counterclockwise right around the point. Even when the mass current does not apparently rotate, the angle of rotation at a fixed point may decrease or increase, as shown in Fig. \ref{fig:zerovorticity}.

\subsection{Teleport of Inverse Mass} \label{sec:teleport_inv}
Let us define a scalar quantity $\sigma$ as
\begin{equation}
    \sigma\equiv\nabla\cdot(\rho\mathbf{v})+\frac{\partial\rho}{\partial t}. \label{sigma}
\end{equation}
Then, $\sigma=0$ means the transport equation of mass of Eq. (\ref{conEq1}). Similarly, we can define $\tilde{\sigma}$ as
\begin{equation}
    \tilde{\sigma}\equiv\tilde{\nabla}\cdot(\tilde{\rho}\tilde{\mathbf{v}})+\frac{\partial\tilde{\rho}}{\partial t}. \label{tsigma}
\end{equation}
Because the velocity is zero from Eq. (\ref{zerov}), we immediately have
\begin{equation}
    \frac{\partial\tilde{\rho}}{\partial t}=\tilde{\sigma} \label{telemass1}
\end{equation}
or, equivalently,
\begin{equation}
    \frac{\partial\log\tilde{\rho}}{\partial t}=\frac{\tilde{\sigma}}{\rho}. \label{telemass1a}
\end{equation}
Similarly, we define $\mathbf{\tau}$ and $\tilde{\mathbf{\tau}}$ as
\begin{equation}
    \mathbf{\tau}\equiv\nabla\times(\rho\mathbf{v})-2\rho\frac{\partial\boldsymbol{\theta}}{\partial t} \label{tau}
\end{equation}
and
\begin{equation}
    \tilde{\mathbf{\tau}}\equiv\tilde{\nabla}\times(\tilde{\rho}\tilde{\mathbf{v}})-2\tilde{\rho}\frac{\partial\tilde{\boldsymbol{\theta}}}{\partial t}, \label{ttau}
\end{equation}
respectively. Then we can express Eq. (\ref{conEq3}) as $\mathbf{\tau}=\mathbf{0}$. Because the velocity of inverse mass is zero from Eq. (\ref{zerov}), we have from Eq. (\ref{ttau})
\begin{equation}
    2\tilde{\rho}\frac{\partial\tilde{\boldsymbol{\theta}}}{\partial t}=-\tilde{\mathbf{\tau}} \label{telemass2}
\end{equation}
or, equivalently,
\begin{equation}
    \frac{\partial(2\tilde{\boldsymbol{\theta}})}{\partial t}=-\frac{\tilde{\mathbf{\tau}}}{\rho}. \label{telemass2a}
\end{equation}
We refer to Eqs. (\ref{telemass1}) and (\ref{telemass2}) or, equivalently, Eqs. (\ref{telemass1a}) and (\ref{telemass2a}) as the teleport equations of inverse mass. $\tilde{\sigma}$ and $\tilde{\mathbf{\tau}}$ are the local rates governing the point-wise dynamics of inverse mass. If we know the velocity of mass density as observational data, we can determine them by
\begin{equation}
    \frac{\tilde{\sigma}}{\tilde{\rho}}=\nabla\cdot\mathbf{v} \label{srcdiv}
\end{equation}
and
\begin{equation}
    \frac{\tilde{\mathbf{\tau}}}{\tilde{\rho}}=\nabla\times\mathbf{v}. \label{srccul}
\end{equation}
We can derive Eqs. (\ref{srcdiv}) and (\ref{srccul}) by
\begin{equation*}
    \frac{\tilde{\sigma}}{\tilde{\rho}}=\frac{\partial\log\tilde{\rho}}{\partial t}=-\frac{d\log\rho}{dt}=\nabla\cdot\mathbf{v}
\end{equation*}
and
\begin{equation*}
    \frac{\tilde{\mathbf{\tau}}}{\tilde{\rho}}=-2\frac{\partial\tilde{\boldsymbol{\theta}}}{\partial t}=2\frac{d\boldsymbol{\theta}}{dt}=\nabla\times\mathbf{v}
\end{equation*}
where we used
\begin{eqnarray}
\frac{\partial\log\tilde{\rho}}{\partial t}+\frac{d\log\rho}{dt}=0 \label{inversion2}\\
\frac{\partial\tilde{\boldsymbol{\theta}}}{\partial t}+\frac{d\boldsymbol{\theta}}{dt}=\mathbf{0} \label{inversion3}
\end{eqnarray}
Eq. (\ref{inversion2}) is derived from Eqs. (\ref{dMVC1_2}), (\ref{dMVC2_2}), (\ref{zerov}), and (\ref{tidt}). Eq. (\ref{inversion3}) is derived from Eqs. (\ref{dthJ3}), (\ref{dthJ4}), and (\ref{tidt}).

The local rates are also connected to the local rate $\tilde{\boldsymbol{\Sigma}}$ for the inverse Jacobian in Eq. (\ref{teleJ1}).
We can take the trace and cross operators to Eq. (\ref{gradv1}) and use Eqs. (\ref{srcdiv}) and (\ref{srccul}) to have
\begin{equation}
    \frac{\tilde{\sigma}}{\tilde{\rho}}=\mathrm{Tr}[\tilde{\boldsymbol{\Sigma}}\tilde{\mathbf{J}}^{-1}]
\end{equation}
and
\begin{equation}
    -\frac{\tilde{\mathbf{\tau}}}{\tilde{\rho}}=2[\tilde{\boldsymbol{\Sigma}}\tilde{\mathbf{J}}^{-1}]_{\times}.
\end{equation}

\begin{figure}[t]
    \centering
    \includegraphics[width=8.8cm]{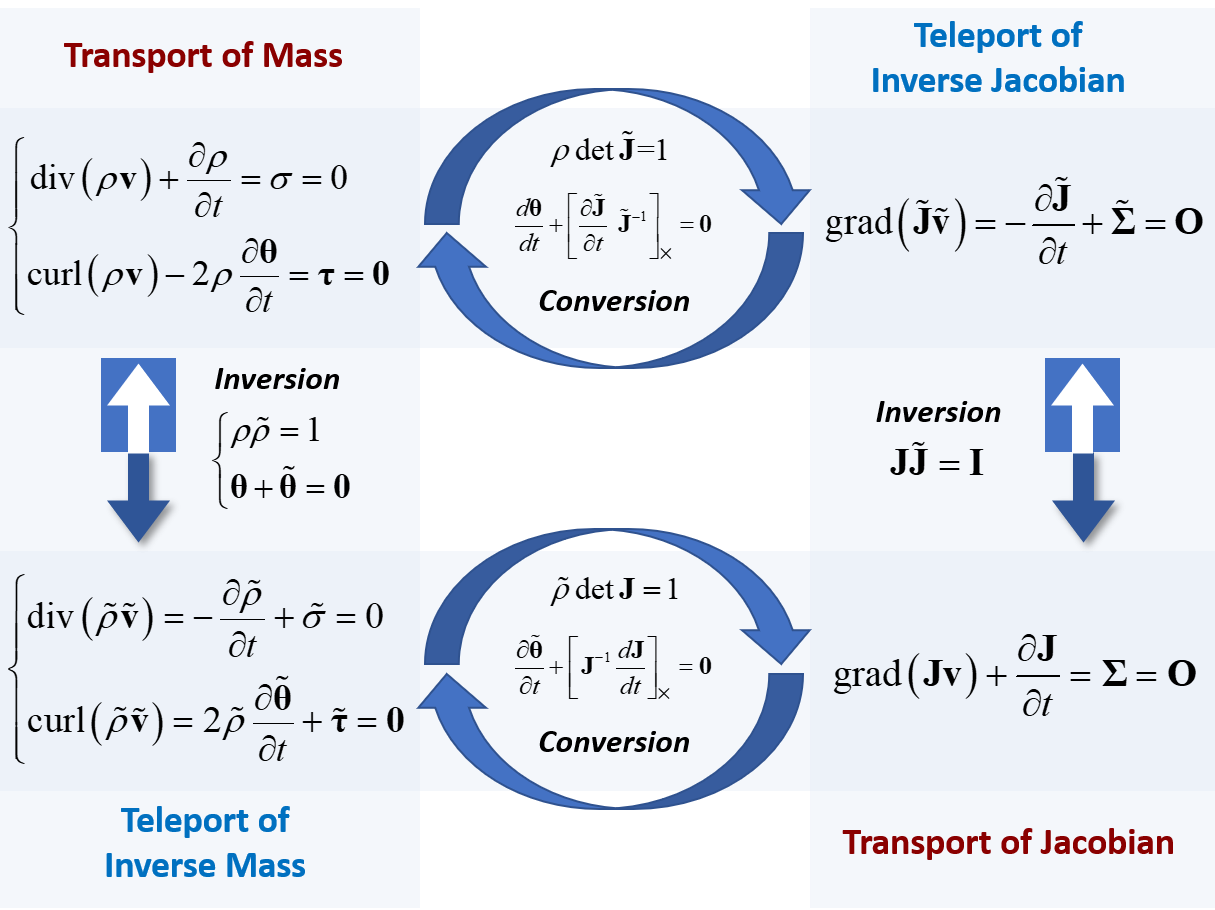}
    \caption{Four different representations of continuum dynamics.}
    \label{fig:Representation}
\end{figure}

Figure \ref{fig:Representation} shows four different representations of continuum dynamics that we have derived in Sections \ref{sec:mass_transport}, \ref{sec:Jacobian_transport}, and \ref{sec:invJacobian_teleport} together with this subsection. The space denoted by tilde symbols is the representation in which machines can efficiently learn physical motions of objects with low complexity models and a smaller number of observational data. In the space, the dynamics of inverse mass density is governed by the teleport equations of inverse mass. Equivalently, its Jacobian is also governed by the teleport equations of Jacobian.

\subsection{Inversion and Viewpoints} \label{sec:viewpoints}
\begin{figure}[htp]
    \centering
    \includegraphics[width=7.5cm]{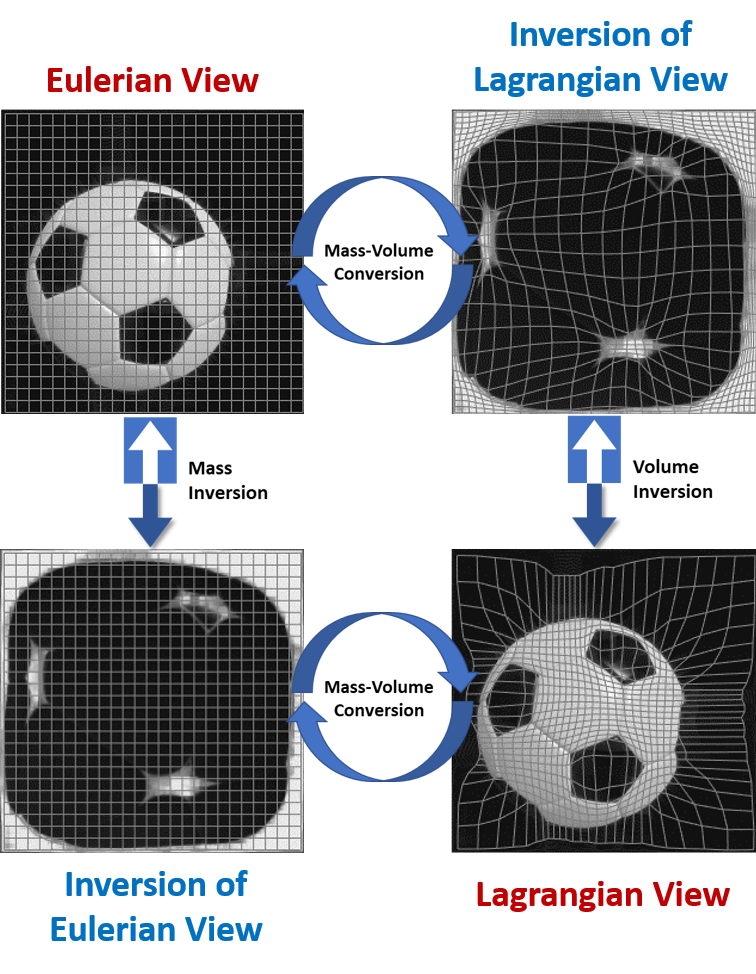}
    \caption{{\bf Two classical viewpoints in fluid mechanics and their inversion}. The inversion views provide a zero velocity representation of the continuum dynamics, which means that we only observe a change of pixel values without motion (\href{https://drive.google.com/file/d/1q1sH4iBms5cA_HTVBKh9VTIIUnfZsX5Y/view?usp=sharing}{click \underline{here} to see a video clip}). The zero velocity representation enables us to model the continuum dynamics of an object with lower complexity and fewer data.}
    \label{fig:view}
\end{figure}

The motion of an object or the flow of a continuum can be described by two traditional viewpoints: Eulerian and Lagrangian. As shown in Fig. \ref{fig:view}, we can revisit them by generalizing the terms of pixels (mass-filled quadrangle meshes) and pixel values (amount of mass per pixel). In the Eulerian view, pixels are fixed squares of unit volume, and pixel values are highly variable even by slight object motion. In the Lagrangian view, pixels move along with material points and have variable sizes and shapes because of the converging, diverging, or whirling motions of material points. Pixel sizes can be selected so that pixel values are all units; multiplying the pixel size by its brightness provides the unit pixel value. Hence, while only pixel values are recorded in the Eulerian view, the shapes and sizes of pixels are required. These are handled by the Jacobian matrices shown in Fig. \ref{fig:Representation}, without the need to record the uniform pixel values in the Lagrangian view.

The viewpoints of the dynamics can be inverted by the inversion transform. The inversion of the Eulerian view is a Eulerian view where it still uses the regular pixel arrays but has a distorted and inverted distribution of mass density. Mass density in the inverted view is still positive but reciprocal to the original density. Similarly, the distorted and inverted distribution is observed in the inversion of the Lagrangian view. However, pixels move with variable sizes and shapes, and pixel values are all units, similar to the Lagrangian view.

The Eulerian view can be transformed to the inversion of the Lagrangian view by mass-volume conversion, which is a mathematical process for converting pixel values of the Eulerian view into pixels sizes of the inverted Lagrangian view. The Lagrangian and inverted Eulerian views can also transform to each other in the same way. Here, we can describe the unusual behavior that differs from observations in the Eulerian viewpoint. When a mass distribution continuously flows, its inverted distribution of mass does not. The inverted mass density gradually increases or decreases in a pointwise manner. The inversion views provide the zero-velocity representation of an object in motion or the flow of a continuum. Notice that the black background of classical views of the soccer ball is not zero intensity but a small positive, which affects the thickness of the white margin in the inversion views.

Figure \ref{fig:view} is a good visualized example that helps understand Fig. \ref{fig:Representation}. The Eulerian view is well-described by the transport equations of mass. The transport equation of Jacobian explains the Lagrangian view. Similarly, inverse views of Eulerian and Lagrangian are described by the teleport equations of inverse mass and inverse Jacobian, respectively.

\subsection{Inversion of Incomplete Observations} \label{sec:inv_obs}
Determining the dynamics of a system from only mass density observations is an ill-posed problem. Complete observations require a pair of mass density and velocity field (or a pair of mass density and local angles of rotation). Direct observations of the velocity field can determine the local rates of Eqs. (\ref{srcdiv}) and (\ref{srccul}), and the dynamics of inverse mass by Eqs. (\ref{telemass1a}) and (\ref{telemass2a}).

We can discretize Eqs. (\ref{telemass1a}) and (\ref{telemass2a}) into $(i,j)$th pixels:
\begin{equation}
\frac{\partial}{\partial t}\log\tilde{\rho}_{ij}\equiv\frac{\partial}{\partial t}\langle\log\tilde{\rho}\rangle_{ij}=\nabla\cdot\langle\mathbf{v}\rangle_{ij}
\end{equation}
and
\begin{equation}
-\frac{\partial}{\partial t}2\boldsymbol{\theta}_{ij}\equiv-\frac{\partial}{\partial t}\langle2\boldsymbol{\theta}\rangle_{ij}=\nabla\times\langle\mathbf{v}\rangle_{ij}
\end{equation}
The operator $\langle\cdot\rangle_{ij}$ means taking the average on the $(i,j)$th pixel. Equivalently, direct observations of velocity also determine the local rates of Eq. (\ref{gradv1}), and the dynamics of inverse Jacobian by Eq. (\ref{teleJ2}). 

In this study, the approach we take for getting inverse mass or inverse Jacobian is to minimize the divergence of the velocity field. We can transform and discretize it by 
\begin{eqnarray}
\int\rho(\nabla\cdot\mathbf{v})^2dV&=&\int\rho(\frac{d\log\rho}{dt})^2dV \nonumber \\
&=&\int(-\frac{\partial\log\tilde{\rho}}{\partial t})^2\rho~dV \nonumber \\
&=&\int(\frac{\partial\log\tilde{\rho}}{\partial t})^2d\tilde{V} \nonumber \\
&\approx&\sum_{ij}(\log\tilde{\rho}_{ij}^{(t+1)}-\log\tilde{\rho}_{ij}^{(t)})^2
\end{eqnarray}
which comprises Eq. (\ref{ErF}). In summary, our problem is stated as

\noindent \textbf{Problem}. {\it When given $T$ observations \{$\rho^{(t)}(x,y)$\} from the Eulerian point of view, find the $m\times n$ array of inverse Jacobian matrices \{$\tilde{\mathbf{J}}^{(t)}_{ij}$\} for $t=1,\ldots,T$ such that
\begin{itemize}
    \item $\sum_{i,j,t=1}^{m,n,T}(\log\tilde{\rho}_{ij}^{(t+1)}-\log\tilde{\rho}_{ij}^{(t)})^2$ be minimized;
    \item $\log|\tilde{\mathbf{J}}_{ij}^{(t)}|+\langle\log\rho^{(t)}\rangle_{ij}=0$;
    \item $(\tilde{\mathbf{J}}_{ij}-\tilde{\mathbf{J}}_{i+1,j+1})\small{\begin{pmatrix}
+1 \\ -1\end{pmatrix}}
=(\tilde{\mathbf{J}}_{i,j+1}-\tilde{\mathbf{J}}_{i+1,j})\small{\begin{pmatrix}
1 \\ 1\end{pmatrix}}$.
\end{itemize}
}

\subsection{Interpolation and Extrapolation} \label{sec:intp}
In this subsection, we propose an interpolation or extrapolation scheme that we can use for modeling the dynamics of inverse Jacobian. We assume that there is a set of the inverse Jacobian data at different times $t = t_1\ldots, t_N$ on all discrete pixels of 2D space from the inverse Eulerian view,
\[\tilde{\mathbf{J}}_{ij}^{(t_1)},\tilde{\mathbf{J}}_{ij}^{(t_2)},\ldots,\tilde{\mathbf{J}}_{ij}^{(t_N)}\]
where $\tilde{\mathbf{J}}_{ij}^{(t_n)}$ is a 2-by-2 inverse Jacobian matrix for the $ij$th pixel at time $t_n$ and already satisfies Eq. (\ref{nocurlJ2D}). We can generally assume a continuous function or model of parameters $\omega$ by
\begin{equation}
    \tilde{\mathbf{J}}_{ij}(t)=\mathbf{F}_{ij}\left(t;\omega\right)
\end{equation}
where $\mathbf{F}_{ij}\left(t_n;\omega\right)=\tilde{\mathbf{J}}_{ij}^{(t_n)}$ for all. The method we took to make the interpolated one satisfy Eq. (\ref{nocurlJ2D}) is the Lagrange interpolating polynomial~\cite{Abramowitz1965}. To write it explicitly, we have
\begin{eqnarray}
\tilde{\mathbf{J}}_{ij}(t)=\frac{(t-t_2)(t-t_3)\cdots(t-t_N)}{(t_1-t_2)(t_1-t_3)\cdots(t_1-t_N)}\tilde{\mathbf{J}}_{ij}^{(t_1)}+\nonumber\\
\cdots+\frac{(t-t_1)(t-t_2)\cdots(t-t_{N-1})}{(t_N-t_1)(t_N-t_2)\cdots(t_N-t_{N-1})}\tilde{\mathbf{J}}_{ij}^{(t_N)}. \label{LagIntp}
\end{eqnarray}
We used this to create the examples shown in Figs. \ref{fig:soccerball}, \ref{fig:cheetah}, and \ref{fig:digit}. An advantage of Lagrange interpolating polynomial is that the weights do not depend on the data. An improved version of the Lagrange formula~\cite{Berrut2004} is also applicable to the problem.

\section{Conclusion}
We proposed a framework for self-exploratory machine learning of transport phenomena.
It is different from many state-of-the-art physics-learned simulators or physics-informed deep learning models because it does not rely on human knowledge (governing equations) or external simulation results as training data (by solving the governing equations). However, it does not reject the effectiveness of deep learning models, and will be combined with them and contribute to the  development  of machine learning models for exploring the physical world.


%




\ifCLASSOPTIONcaptionsoff
  \newpage
\fi

\end{document}